\documentstyle[preprint,revtex]{aps}
\tightenlines
\begin{document}
\draft
\begin{title}
Localized States in Discrete Nonlinear Schr\"{o}dinger Equations
\end{title}
\author{David Cai, ~A.\ R.\ Bishop, and Niels Gr{\o}nbech-Jensen
}
\begin{instit}
Theoretical Division and Center for Nonlinear Studies, Los Alamos National
Laboratory, \\
Los Alamos, New Mexico 87545
\end{instit}
\begin{abstract}
A new 1-D discrete nonlinear Schr\"{o}dinger (NLS) Hamiltonian is
introduced which includes the integrable Ablowitz-Ladik system as a limit.
The symmetry properties of the system are studied.
The relationship between intrinsic localized states and the soliton
of the Ablowitz-Ladik NLS is discussed.
It is pointed out that a staggered localized state
can be viewed as a particle of a {\em negative} effective mass.
It is shown that staggered localized
states can exist in the discrete dark NLS.
The motion of localized states and  Peierls-Nabarro pinning are
studied.
\end{abstract}

\begin{narrowtext}
\narrowtext
Since the theoretical studies  \cite{Sievers,Takeno1,Page} of the
dynamics of a homogeneous nonlinear lattice in
one and higher dimensions (D),
much  attention has been given to  so-called ``intrinsic
localized states'' in a discrete lattice
induced by the nonlinearity of the lattice,
 in contrast to the rather
well understood localized
states induced by point defects in a harmonic lattice.
The properties of the intrinsic localized states have also been
studied with numerical simulations
\cite{Burlakov1,Bourbonnais,Bickham1,Bickham2}
for a simple monatomic lattice of particles with
a nearest-neighbor harmonic and quartic anharmonic interaction
in 1-D and 2-D.
The localized states were found to have amplitude-dependent
frequencies lying above the upper harmonic phonon band edge
and to have their particles oscillating out of phase with their
neighbors.
It has been  demonstrated that the localized states can also
exist in the D-dimensional discrete nonlinear Schr\"{o}dinger (NLS)
lattice \cite{Takeno2}. The work of Ref.\
\cite{Takeno2} has established
that, in a 1-D NLS,
a localized state lying {\em below} the linear phonon band
 reduces to a one-soliton solution in the continuum limit.

In this paper, we propose a new discrete NLS
equation that makes  natural contact with the integrable NLS.
We  discuss
the general properties of the localized states with an emphasis on
the interplay of the integrability and  non-integrability,
the discreteness, and the continuum limit. We will study in detail  a
particular
set of localized solutions that have a staggered form, {\it i.e.},\
the neighboring sites oscillate out of phase. Unlike the unstaggered
localized states, which have the oscillation frequencies below
the linear phonon band and reduce to one-soliton solutions in the
continuum limit, these staggered states have
oscillation frequencies above the  phonon band and have no
continuum counterpart. We will show that
they  can exist even in a discretized NLS
for the dark NLS and that, in particular,
 they can be treated
as particles of {\em negative} effective mass.
These states, staggered or otherwise,
are also reminiscent of ``gap solitons'' which, {\it e.g.},\ give
rise to self-induced
transparency in  electromagnetic wave transmission through a
superlattice whose dielectric constant is periodic in space
\cite{Chen,Mills}. In our case, the underlying
periodicity arises from the intrinsic discreteness of
the system.
Finally, we will analyze the modulational stability of
the system to gain insight into the formation and destruction
of these localized states.
For a  complete picture of our system, we
 also present results from  numerical simulations.

The discrete 1-D NLS equation we study is:
\begin{equation}
i\dot{\phi}_{n}  =  - ( \phi_{n+1}+\phi_{n-1}) \label{indnls}
 - [\mu(\phi_{n+1} + \phi_{n-1})
+ 2 \nu\phi_{n}]|\phi_{n}|^{2},  \nonumber
\end{equation}
where
the overdot stands for the derivative with respect
to time $t$, $n$ is a site index,  and $\mu \geq 0$.
This can be derived from the Hamiltonian:
\[H  =  - \sum_{n} \phi_{n}\phi_{n+1}^{\ast} + \phi_{n}^{\ast}\phi_{n+1}
      \mbox{} - \frac{2\nu}{\mu}\sum_{n}|\phi_{n}|^{2}
 + \frac{2\nu}{\mu^{2}}\sum_{n} \ln(1 + \mu|\phi_{n}|^{2}) \]
with the deformed Poisson brackets
\begin{eqnarray*}
\{\phi_{n},\phi_{m}^{\ast}\} & = & i(1+\mu|\phi_{n}|^{2})\delta_{nm} \\
\{\phi_{n},\phi_{m}\} & = & \{\phi_{n}^{\ast},\phi_{m}^{\ast}\} = 0
\end{eqnarray*}
and the equation of motion
$\dot{\phi}_{n} = \{ H, \phi_{n}\}$.
We will refer to this system as IN-DNLS.
 The system has an
energy conservation law.
The quantity
\[ N = \frac{1}{\mu} \sum_{n}\ln(1 + \mu|\phi_{n}|^{2}) \]
is also conserved and  serves as a norm. Notice that the limits of
 $H, N$ exist, as $\mu \rightarrow 0$.

If $\mu=0$, Eq.\ (\ref{indnls}) reduces to a familiar discrete
NLS equation (referred to as N-DNLS below) which is nonintegrable.
If $\nu = 0$, Eq.\ (\ref{indnls}) is the Ablowitz-Ladik NLS (referred
to as I-DNLS), which is
integrable and possesses an infinite number of
conservation laws\cite{Ablowitz,Faddeev}.
Due to the scaling property between the nonlinear coefficient and the
amplitude, both N-DNLS and I-DNLS have
a single measure for the strength of the nonlinearity, respectively,
 $\nu |\phi|^{2}$ and $\mu|\phi|^{2}$, while IN-DNLS has two,
$\mu|\phi|^{2}$ and $\nu/\mu$.
All these DNLS equations are
 discretizations, up to a trivial gauge transformation, of the integrable
continuum NLS equation, $i\dot{\phi} = - \phi_{xx} - 2\kappa |\phi|^{2} \phi$,
which  possesses bright and dark soliton solutions for
$\kappa > 0$, and $\kappa < 0$, respectively. However,
the discreteness of the systems gives rise to
several interesting features which are not
 present in the continuum limit.

We seek an oscillating solution of IN-DNLS in the form:
\[ \phi_{n} = \psi_{n} e^{-i(\omega t - \alpha n + \sigma_{0}) }, \]
where $\psi_{n}$ is real and $\sigma_{0}$ is a constant phase.
{}From the real and the imaginary parts of Eq.\ (\ref{indnls}), we have
\begin{eqnarray}
(\hat{\Omega}\hat{\psi})_{n} & \equiv &
\omega\psi_{n} + \cos\alpha (\psi_{n+1} + \psi_{n-1}) \label{eigen}\\
                             &        &+ [ \mu \cos \alpha (\psi_{n+1} +
\psi_{n-1})
+ 2\nu \psi_{n}] \psi_{n}^{2} = 0 \nonumber
\end{eqnarray}
\begin{equation}
\dot{\psi}_{n} = - \sin\alpha (\psi_{n+1} - \psi_{n-1})
(1 + \mu \psi_{n}^{2}),
\label{motion}
\end{equation}
where  $\hat{\psi}$ is the column vector,
$\{\psi_{1},\psi_{2},...,\psi_{n}...\}$, and $\hat{\Omega}$ is the matrix
defined by the l.h.s.\ of Eq.\ (\ref{eigen}).
Equation\ (\ref{eigen}) with vanishing boundary condition
constitutes a nonlinear algebraic eigenvalue problem for
localized states.
Equation\ (\ref{motion}) determines the time evolution of the localized
states (Note that the above ansatz and results (\ref{eigen}) and
(\ref{motion}) are readily generalized to ${\rm D} > 1$).
 When $\alpha = 0$ or $\pi$, $\phi_{n}$ is stationary. We shall
call a localized state staggered, if $\alpha = \pi$ and unstaggered,
if $\alpha = 0$. From Eq.\ (\ref{eigen}), we have
\[\omega  =  -2 \cos\alpha
           -\mu\cos\alpha
\frac{\sum_{n}(\psi_{n+1}+\psi_{n-1})\psi_{n}^2}{\sum_{n}\psi_{n}}
-2\nu \frac{\sum_{n}\psi_{n}^{3}}{\sum_{n}\psi_{n}}. \]
Clearly,
if $\psi_{n} > 0$ and $|\nu|$ is not too large, the staggered state lies above
the phonon band, while the unstaggered state lies below.  Particularly,
if $\mu=0$, there is no such localized state, staggered or unstaggered,
below the phonon band for $\nu < 0$, or
above the phonon band for $\nu > 0$. In the
following discussion, we
focus mainly on those staggered
states whose frequencies lie outside the phonon band.

One can easily
show that IN-DNLS possesses the following reflectional symmetry: if
an unstaggered state, $\psi_{n}\exp(-i\omega t)$,
is a solution of the eigenvalue problem (\ref{eigen}),
 then the staggered state,
 $(-1)^{n}\psi_{n}\exp(i\omega t)$, is a solution of the dual eigenvalue
problem,
{\it i.e.}\ Eq.\ (\ref{eigen}) with $\nu \rightarrow -\nu$.
{}From this symmetry,
for N-DNLS, it follows
that there exists a staggered localized state  whose frequency is
above the phonon band for $\nu < 0$ {\it if} there is an unstaggered localized
state below the phonon band for $\nu > 0$. Later we will return to
the stability issue of the staggered localized states in the dark
N-DNLS ({\it i.e.}\ $\mu = 0$, $\nu < 0$).

For I-DNLS, we can exactly solve
the nonlinear eigenvalue problem. The localized solutions are of the form:
\begin{eqnarray}
\phi_{n} & = & \frac{\sinh\beta}{\sqrt{\mu}} {\rm sech}[\beta(n-u t - x_{0})]
e^{-i(\omega t -\alpha n + \sigma_{0})} \label{soliton} \\
\omega~\, & = & \ -2 \cos\alpha \cosh\beta \\
u~\ & = & \ 2 \beta^{-1}\sin\alpha\sinh\beta \label{velocity}
\end{eqnarray}
and have the energy $E = H$:
\begin{equation}
E = -4 \mu^{-1}\cos\alpha\sinh\beta. \label{energy}
\end{equation}
These localized states are precisely the
exact one-soliton solutions obtained via, {\it e.g.},\ the
inverse spectral transform \cite{Ablowitz}.
One can readily show that the solutions in Eq.\ (\ref{soliton}),
under the above reflectional transformation, transform to
a set of solutions identical to
 the original set in Eq.\ (\ref{soliton})
but with a different parametrization. It follows  that
these one-soliton solutions possess this exact self-dual
reflectional symmetry.

Another striking property related to these localized states in I-DNLS is that
they have {\em continuous} translational symmetry and for each $\beta$
there exists a band
of velocities  at which a localized state  can travel
(see Eq.\ (\ref{velocity}))
without experiencing any Peierls-Nabarro (PN) pinning
from the lattice discreteness (see Eq.\ (\ref{energy}) )
\cite{Rainer}. This is
in contrast to the case of N-DNLS in which
a moving localized state experiences dispersion and eventually
decays \cite{Takeno2}.
 We note in passing
that, contrary to the general discrete case,
a soliton  of some fixed amplitude
in the continuum NLS can always be Galileo boosted to any
velocity.

With the nonlinear $\mu$-term, we expect that
the localized states
in IN-DNLS are more robust against dispersion while translating,
 since
the PN barrier is reduced by the presence of the
the nonlinear nearest neighbor interaction. We demonstrate this
by using an ansatz for a localized state
to calculate the total energy $E$ of the state as a function of $\delta$,
the position of the  center of the localized state between two
neighboring lattice sites. The ansatz is
\[ \phi_{n} = (\lambda)^{n}\frac{\beta}{\sqrt{\mu+\lambda \nu}}{\rm
sech}(\beta( n + \delta)) \]
where $\lambda = 1, -1$ for the unstaggered, staggered state, respectively.
This ansatz is a good representation of a localized state from
our analytical study (see  Eq.\ (\ref{ansatz}) below) and numerical
simulation  for small $|\nu|/\mu$ and small $\beta$.
 Notice that, in the terminology of \cite{Page,Bickham2},
such a staggered state
with $\delta = 0.5$ is an even-parity mode, and, $\delta = 0$ is
an odd-parity mode.
This classification loses its usefulness for the solitons in the I-DNLS
which have continuously translational symmetry.
The function $E(\delta)$ shows that, for $\nu < 0 $,
a localized state whose center is located
 at the midpoint between two lattice sites has a
lower total energy than  one located precisely at a lattice site, regardless
of the state being staggered or unstaggered, and for $\nu > 0$, the situation
is reversed (see Fig.\ \ref{fig1}).
The energy difference between $\delta = 0$
and $\delta = 0.5$ becomes more and more pronounced with the increase
of $|\nu|/\mu$ or with the increase of the amplitude of a localized state.
This energy difference will act as a barrier to the
translation of a localized state.
 In our simulation, we found that
a very localized state ({\it i.e.}\ with large amplitude) is pinned
and can not move. By reducing the amplitude or $\nu$, a state
which is very localized
can travel at the {\em group velocity}, $V_{g} \approx 2\sin\alpha$,
 which is  derived from the linear dispersion
relation of the NLS;
this traveling state  continuously emits a small and long phonon tail
and gradually slows down. We also observed that some of the states
of this kind will slow down to a threshold velocity, at which
the state
is abruptly pinned. In this context, we introduce an important concept,
namely that a
staggered localized state can be viewed as a particle
with a {\em negative} effective mass.
{}From Eq.\ (\ref{energy}), for $\alpha = \alpha_{s}$ or
$\pi + \alpha_{s}$, $|\alpha_{s}|\ll 1$,
the energy can be expressed as
\[ E = -\frac{4\lambda\sinh\beta}{\mu}\cos\alpha_{s}
= - \frac{ 4\lambda\sinh\beta}{\mu} + \frac{\lambda\beta^{2}}{2\mu\sinh\beta}
u^{2}, \]
which exhibits  a {\em negative} (positive) effective mass for
 staggered (unstaggered) state. Obviously, a localized state of a
negative effective mass is mechanically unstable (stable) at a
minimum (maximum) of the PN potential and it tends to run
away from the minimum. This is confirmed by our simulation. In Fig.\
\ref{fig1},
we present  cases in which a trapped staggered (unstaggered) localized
state is oscillating at the top (bottom) of the PN potential
for $\nu > 0$ ($\nu < 0$).

Now we turn to discussing localized states in the low
amplitude limit.
As pointed out in \cite{Sievers},  lattice Green's function
methods can be utilized to study the existence of
 nonlinear localized states. From
the linear part of Eq.\ (\ref{eigen}) we derive the lattice Green's
function
\[ G_{\alpha}(n,m) = \frac{1}{N} \sum_{q}
\frac{e^{iq(n-m)}}{\omega_{\alpha}(q)-\omega} \]
\[ \longrightarrow -\frac{({\rm sgn}\omega)^{|n-m| + 1}}
{\sqrt{\omega^{2} - 4 \cos^{2}\alpha}}
{(\frac{\sqrt{\omega^{2} - 4 \cos^{2}\alpha} -
|\omega|}{2\cos\alpha})}^{|n-m|}, \]
as $N \rightarrow \infty$, for $|\omega/(2\cos\alpha)| > 1$,
where $\omega_{\alpha}(q) = - 2 \cos\alpha \cos q$ is the eigenvalue
for the linear part of Eq.\ (\ref{eigen}).
It can be shown that the localized
states satisfy the following equation:
\begin{equation}
\psi_{n}=\sum_{m}G_{\alpha}(n,m)[\mu\cos\alpha(\psi_{m+1}+\psi_{m-1})
+2\nu\psi_{m}]\psi_{m}^{2}. \label{greensolution}
\end{equation}
For a stationary state whose frequency is very close to
the band edge, {\it i.e.}\
 $ \omega = -\lambda (2 + \Delta^2)$, $0 <\Delta \ll 1$,
 we see that
the Green's function has the asymptotic form
\[ G_{\alpha}(n,m) \longrightarrow  \frac{\lambda}{2\Delta}e^{-\Delta|n-m|},
{}~~{\rm as}~~ \Delta \rightarrow 0.\]
Then
Eq.\ (\ref{greensolution}) can be solved
 to obtain an asymptotic localized solution
\begin{equation}
\phi_{n} = (\lambda)^{n}\frac{\Delta}{\sqrt{\mu+\lambda\nu}}
{\rm sech}(\Delta n)e^{i[\lambda(2+\Delta^{2})t]},
\label{ansatz}
\end{equation}
which is a localized state with a large width and a small amplitude.
It can clearly be seen that
a low amplitude staggered state has lower (higher) frequency, $\omega_{IN}$,
than  $\omega_{I}$ of a localized I-DNLS state if they have the same
amplitude and $\nu >0$ ($\nu < 0$).
They are related by
\begin{equation}
\omega_{IN} = \frac{(\mu-\nu)\omega_{I}+2\nu}{\mu}. \label{freq}
\end{equation}
{}From our numerical simulation, we found that
this relationship holds rather well even for very localized staggered
states.(See  Fig.\ \ref{fig2}).

Next we analyze spatially uniform, {\it i.e.}\ n-independent,
 solutions and their stability.
The modulational
instability of these spatially uniform states
is indicative of the numerical stability
of a localized state. As is well known, in the continuum limit,
phonons in bright NLS are linearly unstable and
they will focus to form a soliton,  whereas,  the phonons in dark NLS
are stable and there is no localized solution for vanishing boundary condition.
 From Eq.\ (\ref{eigen}), we have the nonlinear
dispersion relation
\[ \omega = -2\cos\alpha -2 (\mu\cos\alpha + \nu) \psi^{2} \]
for the state $\psi_{n} = \psi$, independent of $n$.
 To study the linear stability of these
states with periodic boundary conditions,
$\phi_{n} = \phi_{n+N}$,
we seek a solution in
the completely general form
\cite{Eilbeck,Carr}:
\[ \phi_{n} = [\psi_{n} + u_{n}(t)] e^{-i(\omega t -\alpha n)}, \]
where $u_{n}$ is a small perturbation.
Defining $\hat{u} = \hat{u}_{1} + i\hat{u}_{2}$,
the linearized equation can be written as
\[ \left( \begin{array}{c}
 \dot{\hat{u}}_{1} \\  \dot{\hat{u}}_{2}
 \end{array} \right) = \left( \begin{array}{cc}
				0 & \hat{\Omega} \\
				\hat{J} & 0
				\end{array} \right)
	\left( \begin{array}{c}
		\hat{u}_{1} \\ \hat{u}_{2} \end{array} \right), \]
where $\hat{\Omega}$ is the matrix defined in Eq.\ (\ref{eigen}),
and $\hat{J}$ is the associated Jacobian matrix.
The solutions are linearly stable if and only if the eigenvalue problem,
\[ \det(\hat{J}\hat{\Omega} - \lambda^{2}\hat{I}) = 0, \]
has only real and non-positive solutions for $\lambda$.
There exists at least one zero
eigenvalue as a consequence of $\hat{\Omega}\hat{\psi} = 0$.
It is straightforward to show
that, for $\mu \geq 0$, the unstaggered spatially uniform solutions are
stable if $ \nu > - \mu \cos^{2} (\pi/N)$ and their amplitude
 $\psi^{2} \leq \sin^{2}(\pi/N)/(\nu + \mu\cos^{2}(\pi/N))$,
and they are alway stable if $\nu \leq - \mu \cos^{2} (\pi/N)$.
The staggered spatially uniform solutions are stable if
 $ \nu <  \mu \cos^{2} (\pi/N)$
and  their amplitude
 $\psi^{2} \leq \sin^{2}(\pi/N)/(\mu\cos^{2}(\pi/N)-\nu)$, and are always
stable if $ \nu \geq \mu \cos^{2} (\pi/N)$. From the stability of these
 solutions we expect that, for $\nu < \mu$,
the staggered localized states are stable since
the staggered phonons
are generally unstable as $N \rightarrow \infty$, and,
 in  particular,
the dark N-DNLS
has  stable staggered localized states
and the unstaggered localized states decay to stable unstaggered
phonons. These phenomena were indeed observed in our simulation. For
example,
an initially staggered localized state re-adjusts its shape to
become a stable localized state,
while an initially unstaggered localized state
 always decays to a spatially
extended, unstaggered small amplitude state for the dark N-DNLS.

In conclusion,
the new discrete NLS we proposed here has
enabled us to demonstrate clearly how the reflection symmetry and translational
symmetry of the integrable DNLS are broken by on-site nonlinearity. We
have pointed out that the localized states in I-DNLS in the sense of
\cite{Sievers,Takeno1,Page} are the Ablowitz-Ladik solitons. We
have demonstrated that the motion of a staggered state can be understood as
a particle of  negative effective mass.
We have also shown that staggered localized states exist and are stable
 in the dark N-DNLS.
Furthermore,
the analysis of the modulational instability of spatially uniform states
has deepened our understanding of the creation and decay processes of a
localized
state. We
have also discussed the properties of Peierls-Nabarro potential which is
intimately related to the non-integrability of the Hamiltonian and which
can be continuously tuned in our model.
 This controlling mechanism can be utilized in the study of the transport
properties of the localized states, especially in the study of the
{\em dynamical} competition of the localizations induced by nonlinearity and
by randomness in the Anderson sense.

We thank F. Bronold for interesting discussions and
 acknowledge support for this work by the U.S. D.o.E.

\pagebreak

\figure{\label{fig1}
Time evolution of the amplitudes $|\phi_{n}|$ at $n = 100$\ (thin line),
$101$\ (thick line) are shown here
for (a) a trapped staggered localized state
that oscillates around $\delta = 0.5$ at the top of
the Peierls-Nabarro\ (PN) potential for $\nu = 0.1$ with
the initial parameters $\alpha = 0.314 \times 10^{1}$,
$\beta = 1$, $x_{0} = 100.5$;
(b) an unstaggered localized state at the bottom of the
PN potential for $\nu = - 0.1$
with the initial parameters $\alpha = 0.2 \times 10^{-3}$,
$\beta = 1$, $x_{0} = 100.5$.
The insets show their corresponding PN potentials ($\mu = 1$) (see text).}
\figure{\label{fig2}
``$\times$'' shows frequency, $\omega_{IN}$, obtained by Fourier
transform, of a staggered
localized state in IN-DNLS\ ($\mu=1$);
as reference, ``$\Box$'' shows the frequency, $\omega_{I}$,
of the localized state of the same amplitude in I-DNLS\ ($\mu=1$).
``$+$'' shows $\omega_{IN}$ predicted by Eq.\ (\ref{freq}). The
higher branch is for states with $\beta \approx 1$ and the
lower one is for less localized states with $\beta \approx 0.5$
(see text).
}

\end{narrowtext}
\end{document}